\documentclass[aps, prd, superscriptaddress, nofootinbib, twocolumn, preprintnumbers]{revtex4-1}

\usepackage{bm,amssymb,slashed,graphicx,multirow,soul,mathtools,xspace,array}
\usepackage{enumitem}
\usepackage{makecell}
\usepackage{xcolor}

\definecolor{nicered}{rgb}{0.7,0.1,0.1}
\definecolor{nicegreen}{rgb}{0.1,0.5,0.1}
\definecolor{niceblue}{rgb}{0.1,0.1,0.5}

\usepackage[bookmarks=false]{hyperref}
\hypersetup{colorlinks,citecolor= nicegreen,linkcolor= niceblue,urlcolor= niceblue}

\newcommand{\nn}{\nonumber}

\newcommand{\Kx}{{K^{(*)}}}
\newcommand{\Ks}{{K^{*}}}

\makeatletter
\g@addto@macro\bfseries{\boldmath}
\makeatother

\allowdisplaybreaks
\begin{document}

\title{Lepton universality violation from neutral pion decays in $R_{K^{(*)}}$ measurements}

\author{Dean J.\ Robinson}
\affiliation{Ernest Orlando Lawrence Berkeley National Laboratory, 
University of California, Berkeley, CA 94720, USA}

\begin{abstract}
I show that the neutral pion decay in $B \to K^{(*)} \pi^0 \gamma$, with $\pi^0 \to ee \gamma$, 
might generate large sources of lepton flavor universality violation (LFUV) in measurements of the ratios, $R_{K^{(*)}}$:
If the photons in the $K^{(*)} e e \gamma \gamma$ final state are reconstructed as Bremsstrahlung,
the recovered electron-positron invariant mass can be pushed into the $1$--$6$\,GeV$^{2}$ signal region,
artificially enhancing the measured $B \to K^{(*)} e e$ branching ratio compared to $B \to K^{(*)} \mu\mu$.
I present a conservative estimate and simulation of the $B \to K \pi^0 \gamma$ LFUV background at LHCb, 
that together suggest this effect could reduce the recovered $R_K$ up to several percent.
A reliable assessment of the size of this effect will require dedicated simulations within experimental frameworks themselves. 
\end{abstract}

\maketitle

\section{Introduction}

Measurements of lepton flavor universality violation (LFUV) involving charged dilepton final states are long-known to exhibit LFUV from virtual photon poles.
In a generic semileptonic process $X \to (\gamma^* \to \ell \ell) Y$,
the branching ratio of light versus heavier lepton pair production is enhanced whenever the virtual photon momentum, $q$, 
falls below the pair-production threshold of the heavier pair,
because of the $1/q^2$ pole from the virtual photon exchange in the amplitude.
In the context of $R_\Kx$ measurements, in which the LFUV ratio
\begin{equation}
	R_\Kx = \frac{\text{Br}[B \to \Kx \mu\mu]}{\text{Br}[B \to \Kx ee]}\,,
\end{equation}
the virtual photon exchange in $B \to \Ks(\gamma^* \to \ell\ell)$ leads to significant LFUV in $R_\Ks$ if $q^2 \ll 4m_\mu^2$.

Another effect of this type can arise in $R_\Kx$ from $B \to \eta' \Kx$, 
with the subsequent Dalitz decay $\eta' \to \ell \ell \gamma$~(see e.g. Ref.~\cite{Bordone:2016gaq}).
The photon pole in the $\eta'$ decay leads to approximately a factor of $4$ enhancement of the $\eta' \to ee\gamma$ branching ratio versus $\eta' \to \mu\mu\gamma$:
$\text{Br}[\eta' \to ee \gamma] = 4.91(27) \times 10^{-4}$ versus $\text{Br}[\eta' \to \mu\mu \gamma] = 1.13(28) \times 10^{-4}$~\cite{Zyla:2020zbs}.
This $B \to (\eta' \to \ell\ell \gamma)\Kx$ cascade may then enhance the measured $B \to \Kx ee$ branching ratio,
whenever the final state photon is close enough to the electron or positron to be misreconstructed as Bremsstrahlung, or soft enough to be missed.
At LHCb in particular, the relatively large boost of the $\eta'$ system suggests one could naively expect an $\mathcal{O}(1)$ fraction of such photons to look like Bremsstrahlung:
Noting the branching ratio $\text{Br}[B \to \eta' K] = 7.04(25)\times 10^{-5}$, 
this $B \to \eta' K$ LFUV background is known to lead to a percent level correction to $R_K$~\cite{Bordone:2016gaq}.
For this background, however, the misreconstructed dilepton invariant mass, $q^2_{\text{reco}} \simeq m_{\eta'}^2 < 1$\,GeV$^{2}$, 
and therefore it is only relevant to the signal region with dilepton invariant mass $q^2 < 1$\,GeV$^{2}$.
In the signal region $1.1 < q^2 < 6$\,GeV$^2$, LHCb has recently measured $R_{K^+} = 0.846^{+0.044}_{-0.041}$~\cite{LHCb:2021trn,LHCb:2019hip}
(see also the very recent Ref.~\cite{LHCb:2021lvy}),
in notable tension with the SM prediction $1.00 \pm 0.01$~\cite{Bobeth:2007dw,Descotes-Genon:2015uva,Bordone:2016gaq}  
(cf. Ref.~\cite{Isidori:2020acz}).

\begin{figure}[t]
	\includegraphics[width = 4.5cm]{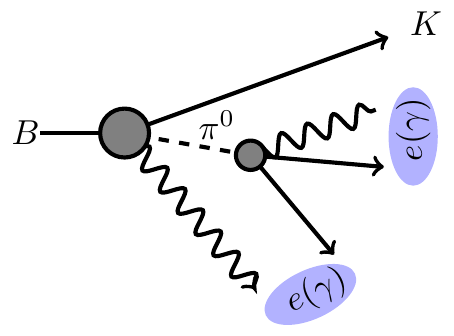}
	\caption{Schematic configuration of $B \to K (\pi^0 \to ee \gamma)\gamma$, with both photons misreconstructed as Bremsstrahlung.}
	\label{fig:misreco}
\end{figure}

In this note, I point out that $b \to s\gamma$ final states involving neutral pion Dalitz decays 
may produce an additional significant SM source of LFUV in $R_{\Kx}$ measurements.
This has not been considered---or at least, not mentioned---as 
a systematic uncertainty in prior LHCb~\cite{LHCb:2014vgu,LHCb:2017avl,LHCb:2019hip,LHCb:2021trn,LHCb:2021lvy}, 
Belle~\cite{Belle:2009zue,Belle:2019oag,BELLE:2019xld} or BaBar~\cite{BaBar:2008jdv} analyses.
In particular, I focus on the $B \to K (\pi^0 \to ee \gamma) \gamma$ LFUV background at LHCb,
as a representative of a possibly broader class of decays with the following pathology:
The photon produced by the $b \to s \gamma$ transition could, on rare occasion, 
be reconstructed into the $\pi^0$ daughter electron or positron as Bremsstrahlung,
so that the misreconstructed $q^2$ of the electron-positron pair may be pushed up into the $1 < q^2 < 6$\,GeV$^{2}$ signal region.
This enhances the measured branching fraction for $B \to K ee$ in the signal region and thus reduces $R_K$.
A schematic of the misreconstruction configuration of the $\pi^0$ decay is shown in Fig.~\ref{fig:misreco},
in which both photons reconstruct as Bremsstrahlung, creating a LFUV background.
Other $b \to s \gamma$ transitions might also contribute similarly to $R_K$, 
including $B \to K h^0 \gamma$ for any neutral (pseudo)scalar meson $h^0 = \eta, \eta', \ldots$,
as well as modes such as $B \to K \pi^0\pi^0\gamma$.
The same effect may occur in $R_{K^*}$ measurements.

The $e^+ e^- \gamma$ invariant mass could be measured precise enough at LHCb---nominally
at $\mathcal{O}(10\,\text{MeV})$ uncertainty~\cite{LHCb:2021rou}---to
reconstruct the $\pi^0$,
providing a handle to reject this background.
However, combinatoric challenges plus related studies with merged photons~\cite{CalvoGomez:2042173} 
suggest this cannot be done with the high efficiencies required,
especially in the case that the daughter photon of the $\pi^0$ may itself either be reconstructed as Bremsstrahlung 
or is soft enough to be lost.
Because the neutral pion is highly boosted at LHCb, the former scenario itself is not expected to be rare---we 
show below it occurs for approximately half of the relevant $\pi^0$ decays---while
the latter scenario is infrequent.

How big could such a LFUV background for $R_K$ be? 
Naively, one expects the leading contribution to $B \to K \pi^0 \gamma$ from the resonant channel $B \to K^* \gamma$.
Since $\text{Br}[B^+ \to (K^{*+} \to K^+\pi^0) \gamma] \simeq 1/3\times3.92(22) \times 10^{-5}$ and $\text{Br}[\pi^0 \to ee \gamma] = 1.174(35)\times 10^{-2}$~\cite{Zyla:2020zbs}, 
the branching ratio for $B \to K ee \gamma \gamma$ is comparable to that of $B \to K ee$ for $1 < q^2 < 6$\,GeV$^{2}$. 
Thus, even if only a few percent of $B \to K ee \gamma \gamma$ were to be misreconstructed as $B \to K ee$, 
a comparable reduction arises in the recovered $R_{K}^{\text{rec}}$,
comparable to the tension with the SM seen at LHCb.

In this note I explore how to estimate the size of this effect,
and then develop an approximate simulation based on a rough, but conservative, 
guesstimate for the implementation of the LHCb upstream Bremsstrahlung recovery algorithm.
This simulation suggests an LFUV background present 
up to the several percent level in $R_K$ from $B \to K \pi^0 \gamma$ alone,
but of course subject to sizeable uncertainties inherent to such approximations.

\section{Differential Branching Ratio}
\subsection{$B \to K \pi^0 \gamma$}
To estimate the branching ratio for $B \to K(\pi^0 \to ee\gamma)\gamma$,
one must first determine the $B \to K \pi^0 \gamma$ amplitude.
To do so, since we are interested only in an estimate,
I approximate this amplitude by the resonant contributions from the vector meson exchange, $B \to (V \to K \pi^0) \gamma$, 
in which $V$ is any strange vector meson with the appropriate $J^{P} = 1^{-}$ quantum numbers. 
The first few such known mesons are shown in Table~\ref{tab:svm}.

\begin{table}[t]
\renewcommand*{\arraystretch}{1.6}
\newcolumntype{C}{ >{\centering\arraybackslash $} m{2.2cm} <{$}}
\newcolumntype{D}{ >{\centering\arraybackslash $} m{1.8cm} <{$}}
\newcolumntype{L}{ >{\raggedright\arraybackslash $} m{1.5cm} <{$}}
\scalebox{0.84}{\parbox{1.2\linewidth}{
\begin{tabular}{LDDCC}
	\hline\hline
	\text{Meson (V)} & \text{Mass [GeV]} & \text{Width [GeV]} & \text{Br}[B^+ \to V^+ \gamma] & \text{Br}[V \to K \pi^0]\\
	\hline
	K^*	 &  0.8917(2) & 0.0514(8) & 3.92(22) \times 10^{-5} & \simeq1/3 \\
	K^*(1410) & 1.414(15) & 0.232(21) & 2.7^{(+0.8)}_{(-0.6)} \times 10^{-5} & 2.2(4) \% \\
	K^*(1680) & 1.718(18) & 0.322(110) & 6.7^{(+1.7)}_{(-1.4)} \times 10^{-5} & 12.9(8)\% \\
	\hline\hline
\end{tabular}
}}
\caption{Data for strange vector mesons with $J^P = 1^-$~\cite{Zyla:2020zbs}. Isospin is assumed to determine $K \pi^+$ versus $K \pi^0$ branching ratios.}
\label{tab:svm}
\end{table}

\begin{figure}[t]
	\includegraphics[width = 3cm]{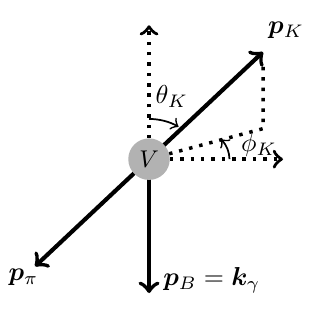}\hfil
	\includegraphics[width = 3cm]{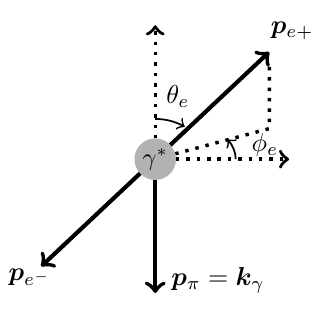}\
	\caption{Left: Definition of the polar helicity angles $\theta_K$ and $\phi_K$ in the $V$ rest frame, for $B \to (V \to K\pi)\gamma$. 
	Right: Definition of the polar helicity angles $\theta_e$ and $\phi_e$ in the virtual photon rest frame, for $\pi^0 \to (\gamma^* \to ee)\gamma$.}
	\label{fig:polar}
\end{figure}

Assuming short-distance dominance (operator $\mathcal{O}_7$), the effective operator mediating the exclusive $B \to V \gamma$ decay 
takes the form $\epsilon^{\mu\nu\rho\sigma} B F_{\rho \sigma }\partial_\mu V_\nu$, in which $F_{\rho\sigma}$ is the photon field strength.
The corresponding amplitude $A_{\lambda \kappa}[B \to V \gamma] \sim (-1)^\lambda\delta^{\lambda\kappa}\,|\bm{k}_\gamma| m_B$,
 in which $\lambda = \pm1,0$ ($\kappa = \pm1$) is the spin (helicity) of the vector meson (photon) in the helicity basis.
The subsequent $V \to K \pi$ amplitude is simply the usual spin-$1$ spherical harmonic, $A_\lambda[V \to K \pi] \sim |\bm{p}_K| d^1_{\lambda 0}(\theta_K, \phi_K)$.
Here $|\bm{p}_K|$ ($|\bm{k}_\gamma|$) is the $K$ ($\gamma$) momentum in the $V$ ($B$) rest frame, and the helicity angles $\theta_K$ and $\phi_K$ are defined in Fig~\ref{fig:polar}.
Because the $\pi^0$ is spin-$0$, $\phi_K$ will be unphysical in the $B \to K(\pi^0 \to ee\gamma)\gamma$ cascade.

We will be interested only in the normalized differential rates, scaled by the appropriate overall branching ratios. 
I have therefore dropped form factors and other normalizing factors both above and hereafter.
The \emph{differential branching ratio} can be shown to take the simple form
\begin{equation}
	\label{eqn:Bdiffbr}
	\frac{d \text{Br}[B \to (V \to K \pi^0) \gamma]}{d \cos\theta_K ds} = \frac{3}{4\pi} |F(s)|^2 \sin^2\theta_K \,,
\end{equation}
with the amplitude summed over resonances,
\begin{align}
	& F(s)  = \sum_V \Bigg\{ \Big[\text{Br}[B \to V \gamma] \text{Br}[V \to K \pi^0]\Big]^{1/2}  \label{eqn:Fs}\\
	& \times \bigg[\frac{ |\bm{p}_K(s)| |\bm{k}_\gamma(s)|}{|\bm{p}_K(m_V^2)| |\bm{k}_\gamma(m_V^2)|}\bigg]^{3/2} 
	\frac{\sqrt{m_V}}{s^{1/4}} \frac{ \sqrt{m_V\Gamma_V}}{s - m_V^2+ i m_V \Gamma_V}\Bigg\}\,. \nn
\end{align}
Here, $s \in [(m_K + m_\pi)^2, m_B^2]$ denotes the virtual $V$ invariant mass, i.e. $s = p_V^2$,
while $|\bm{p}_K(s)| = \sqrt{(s + m_K^2 - m_\pi^2)^2/(4s) - m_K^2}$ and $|\bm{k}_\gamma(s)| = (m_B^2 - s)/(2m_B)$.
Note that in the narrow width limit for a single resonance, $\int ds |F(s)|^2 = \pi \times \text{Br}[B \to V \gamma] \text{Br}[V \to K \pi]$.
In Eq.~\eqref{eqn:Fs}, a simple Breit-Wigner parameterization for the vector meson resonances is used, 
which is appropriate since they lie well above the $K\pi$ threshold,
i.e., $[m_V-(m_K +m_\pi)]/\Gamma_V \gg 1$ 
(see Chapter 49 of Ref.~\cite{Zyla:2020zbs}; see also the discussion of more refined methods therein).

\subsection{$\pi^0 \to ee \gamma$}

Apart from $B \to K(\pi^0 \to ee\gamma)\gamma$, the $Kee\gamma\gamma$ final state may also be accessed from the virtual photon/$Z$ process
$B \to K (\pi^0 \to \gamma\gamma) (\gamma^*/Z^* \to ee)$.
These contributions may be of comparable size, but the interference between them is expected to be negligible, 
because in the former the $ee\gamma$ invariant mass is constrained to the narrow $\pi^0$ resonance, while in the latter this constraint applies to the diphoton mass.
Thus we shall consider only the $B \to K(\pi^0 \to ee\gamma)\gamma$ cascade, 
keeping in mind that our final result will likely be an underestimate of the full $B \to K \pi^0 \gamma \to K ee\gamma\gamma$ rate.
From the most conservative perspective, 
because the photons in $\pi^0 \to \gamma\gamma$ must have the same helicity, 
the virtual $\gamma^*$/$Z^*$ contribution cannot interfere with the contribution from $B \to K(\pi^0 \to ee\gamma)\gamma$ that has opposite helicity photons, 
which accounts for half the total rate.
Thus, if there happened to be fully destructive interference, it would at most reduce our estimate by a factor of two.

The differential rate for $\pi^0 \to ee\gamma$, 
neglecting subleading radiative corrections in this discussion,
\begin{multline}
	\label{eqn:pidiffrate}
	\frac{d \Gamma[\pi \to ee \gamma]}{d \cos\theta_e d s'} = \frac{\alpha \sqrt{s' - 4m_e^2}(m_\pi^2 -s')^3}{512 \pi^2 f^2 m_\pi^3 {s'}^{5/2}} \\ \times \Big[s'\big(\cos 2\theta_e + 3\big) + 8 m_e^2 \sin^2\theta_e\Big]\,,	
\end{multline}
in which $f$ is the effective decay constant of the $(\pi/f) F \tilde{F}$ operator, $s' \in [4m_e^2, m_\pi^2]$ is the electron-positron invariant mass, and $\theta_e$ is the helicity angle defined in Fig~\ref{fig:polar}.
Straightforward integration of the differential rate \eqref{eqn:pidiffrate} allows one to determine the total rate, and hence the normalized $\pi \to ee\gamma$ rate $(1/\Gamma) \, d \Gamma/d \cos\theta_e d s'$.

In the full cascade $B \to (V \to K(\pi^0 \to ee\gamma))\gamma$, taking the narrow width approximation for the $\pi^0$ resonance, 
the full differential branching ratio is then composed as
\begin{align}
	&\frac{d \text{Br}[B \to (V \to K(\pi^0 \to ee\gamma))\gamma]}{d \cos\theta_K\, ds \, d \cos\theta_e\, ds'} \label{eqn:fulldiff}\\
	&= \frac{d \text{Br}[B \to (V \to K \pi^0) \gamma]}{d \cos\theta_K\, ds} 
	\frac{\text{Br}[\pi \to ee \gamma]}{\Gamma[\pi \to ee \gamma]} \frac{d \Gamma[\pi \to ee \gamma]}{d \cos\theta_e\, d s'}\,. \nn
\end{align}
When combined with Eqs.~\eqref{eqn:Bdiffbr} and~\eqref{eqn:pidiffrate}, 
one may then determine the appropriate differential branching ratio weight of any kinematic configuration in the full cascade.

\section{Simulation}

The goal is to simulate the differential distribution 
of the (mis)reconstructed dilepton invariant mass $q^2_{\text{reco}}$, 
while imposing appropriate cuts and requirements for the two photons to reconstruct as Bremsstrahlung in the lab frame.
This can be defined equivalently as
\begin{align}
	 & q^2_{\text{reco}} \equiv (p_B - p_K)^2 \\ & = m_B^2 + m_K^2 - \frac{E_K(m_B^2 + s) + 2 m_B |\bm{p}_K| |\bm{k}_\gamma| \cos\theta_K}{\sqrt{s}}\,, \nn
\end{align}
with $E_K$ the energy of the kaon in the $V$ rest frame.

I simulate the $q^2_{\text{reco}}$ differential distribution via a combination of unweighted Monte Carlo (MC) samples, 
as described below, and differential reweighting.
In particular, with the exception of the $B$ boost distribution and the $K\pi$ invariant mass, $s$, 
a pure phase sample of the full cascade is created,
and then reweighted according to the differential branching ratio weight~\eqref{eqn:fulldiff}. 
The reweighted sample is then binned according to the desired observable---i.e. $q^2_{\text{reco}}$---imposing 
lab frame reconstruction or cut requirements.

\subsection{$B$ boost and $K\pi$ resonances}
\begin{figure}[t]
	\includegraphics[width = 7cm]{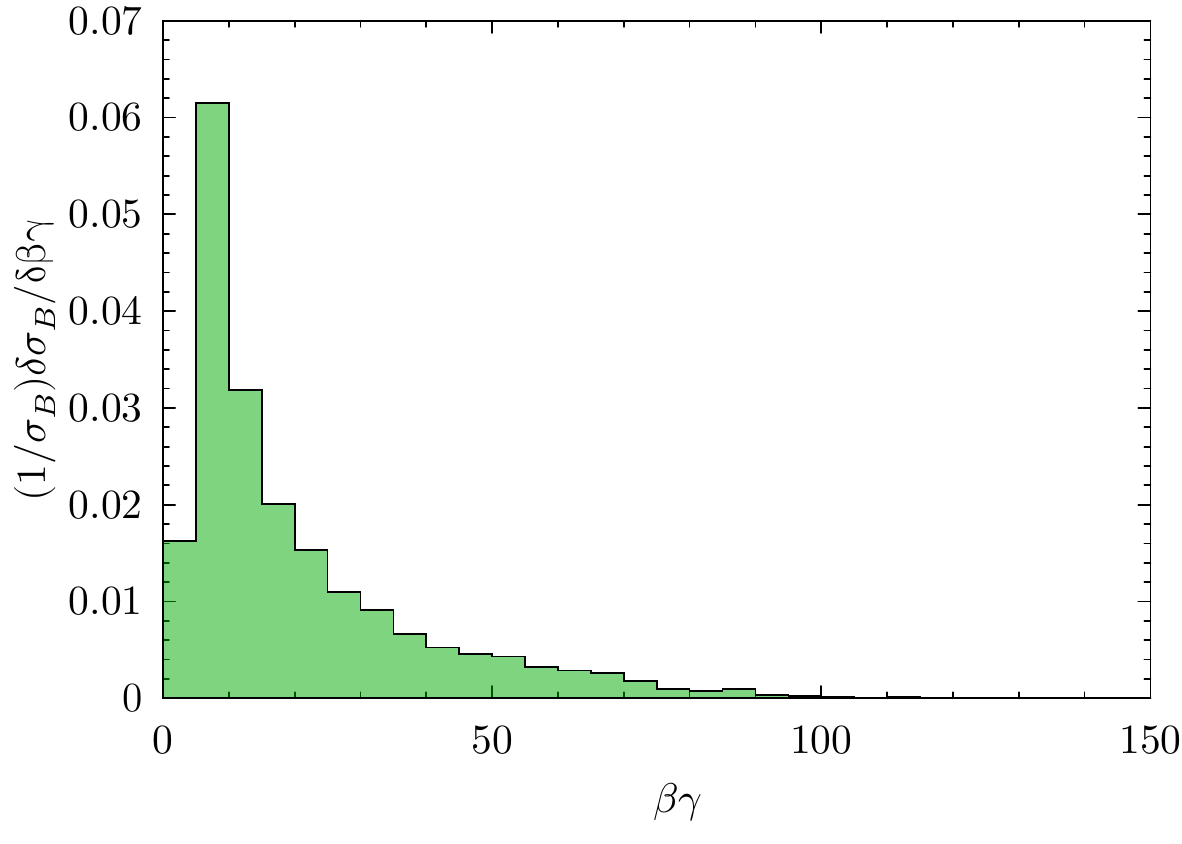}
	\caption{Differential production cross-section for $B$ mesons in the LHCb acceptance with respect to the $B$ boost, $\beta\gamma$.}
	\label{fig:Bboost}
\end{figure}

Simulation of lab frame observables requires sampling the $B$ meson boost distribution in the LHCb acceptance, 
against which the differential weights from Eq.~\eqref{eqn:fulldiff} must be composed.
Simulation of $B$ meson production is done with \texttt{Pythia 8}, 
enforcing $p_T \ge 5$\,GeV and requiring the pseudorapidity $2 \le \eta \le 5$. 
The resulting $B$ boost distribution of the simulated sample is shown in Fig.~\ref{fig:Bboost}. 
The mean boost in this sample is $\langle \beta \gamma \rangle \simeq 20.5$, 
which is very close to quoted averages~\cite{Aaij:2016mos}.

\begin{figure}[t]
	\includegraphics[width = 7cm]{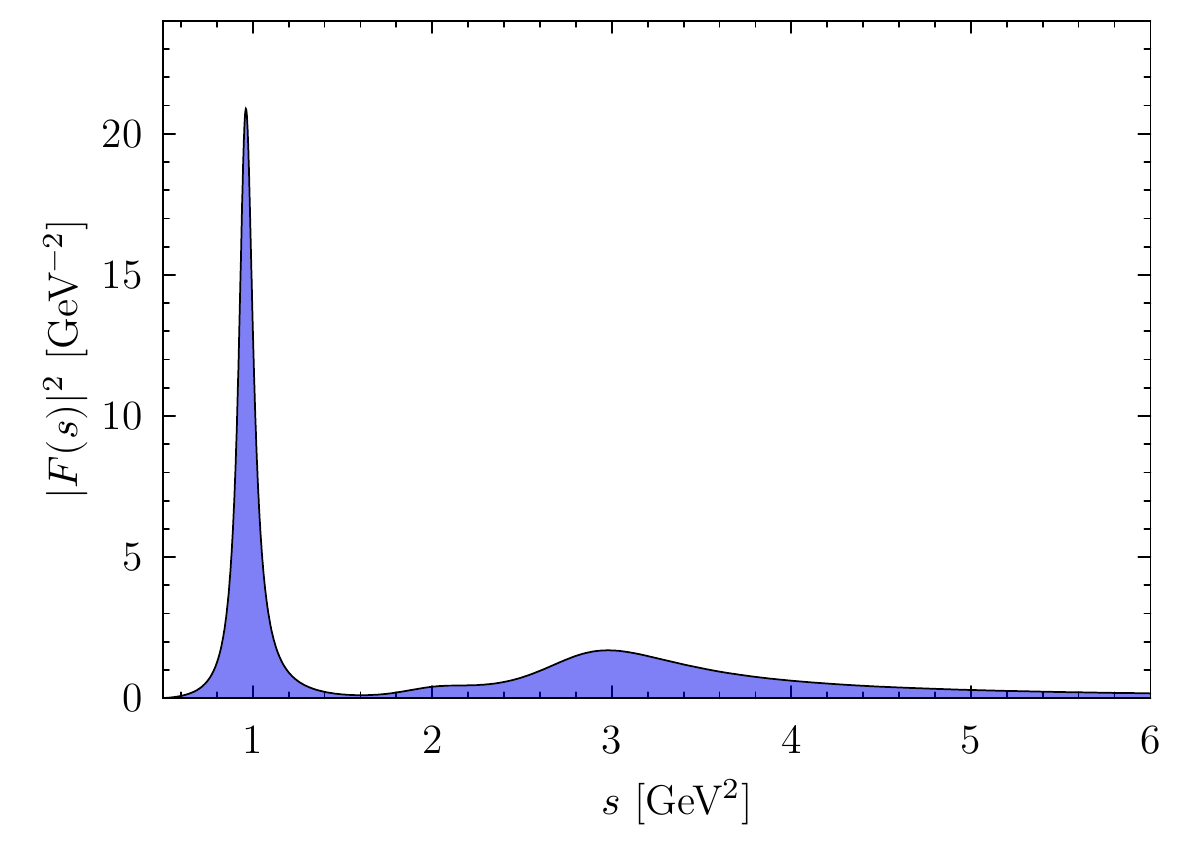}
	\caption{Square amplitude, $|F(s)|^2$, for the three vector meson resonances in Table~\ref{tab:svm}.}
	\label{fig:BWdist}
\end{figure}

From the data in Table~\ref{tab:svm}, the square amplitude $|F(s)|^2$ is shown in Fig.~\ref{fig:BWdist}. 
When further normalized against $K^*$ resonance contribution, 
the integral $\int ds |F(s)|^2/\pi \simeq 1.95\times \text{Br}[B \to K^* \gamma] \times \text{Br}[K^* \to K \pi^0]$, 
so that approximately half of the contribution to the branching ratio comes from the two higher resonances.
Rather than reweighting from a uniform distribution in $s$, the reasonably sharp peak at the $K^*$ resonance
makes it more efficient to create an unweighted sample of the distribution of the $K\pi$ invariant mass, $s$.
For this purpose I create a sample of $10^5$ events.

\subsection{Bremsstrahlung Recovery}
\label{sec:brems}
\begin{figure}[t]
	\includegraphics[width = 0.9\linewidth]{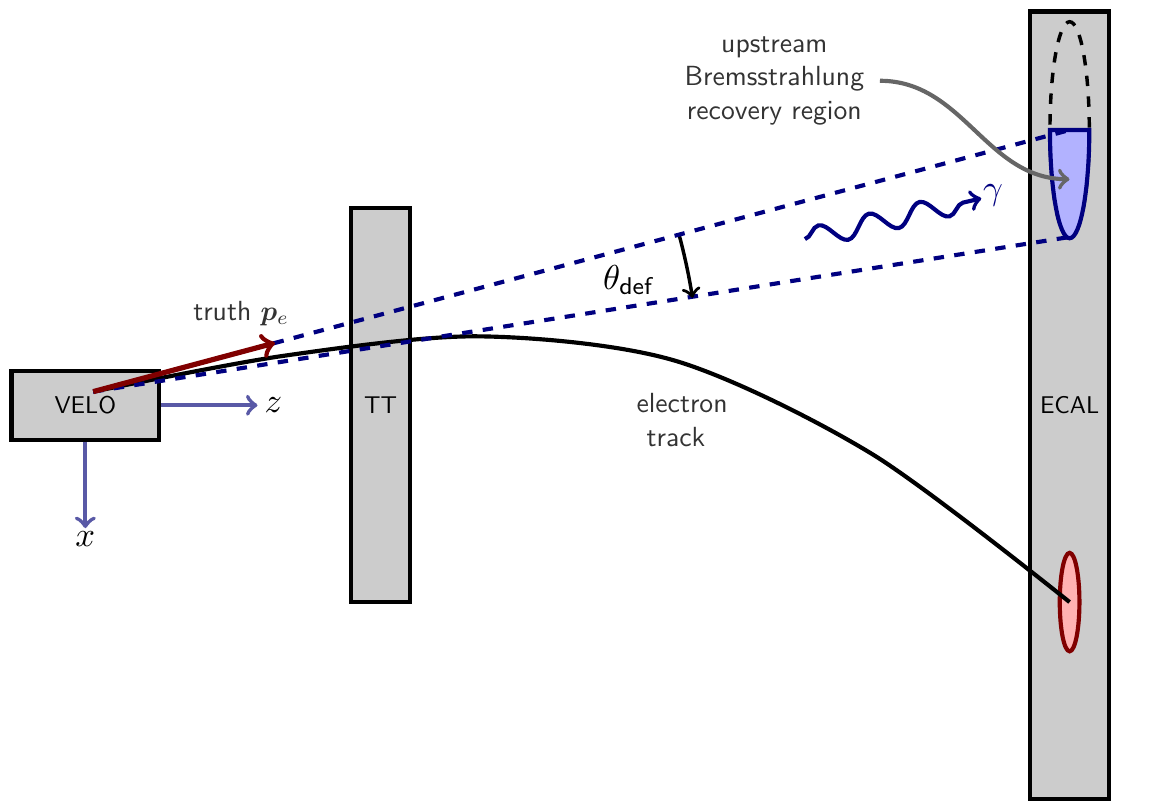}
	\caption{Schematic for upstream Bremsstrahlung recovery at LHCb. 
	Adapted with permission from Ref.~\cite{Berninghoff:2146447}, 
	as reproduced in Fig.~4.4 of Ref.~\cite{Mombacher:2020jrx}.}
	\label{fig:brems}
\end{figure}

The details of Bremsstrahlung photon recovery within the LHCb analysis framework are not available to those external to the collaboration.
One may find, however, approximate or rough details provided in various conference notes and public theses.
For instance, Ref.~\cite{Aguilo:1000431} provides a (possibly somewhat dated) 
study of the recovery of radiation lost by leptons in $B \to J/\psi(ee) K_S$. 
Figure 4.4 of Ref.~\cite{Mombacher:2020jrx} 
(as reproduced from Ref.~\cite{Berninghoff:2146447}; see also Fig.~1 of Ref.~\cite{Aguilo:1000431}), 
from which Fig.~\ref{fig:brems} is adapted,
provides perhaps the clearest visual clue to the Bremsstrahlung recovery:
Photons whose deposit into the ECAL lies within the extrapolated angular 
deflection of a lepton track as it bends from the LHCb VELO through the TT to the magnet---the 
bending of an upstream or a long track---are 
considered compatible with upstream Bremsstrahlung.
The precise region over which photon emission is considered compatible with upstream Bremsstrahlung 
is not specified in available literature,
but instead characterized as the region `before the magnet'~\cite{Mombacher:2020jrx,Aguilo:1000431},
or before the `region with sizeable magnetic field'~\cite{Terrier:691743}.

\begin{figure}[!t]
	\includegraphics[width = 8cm]{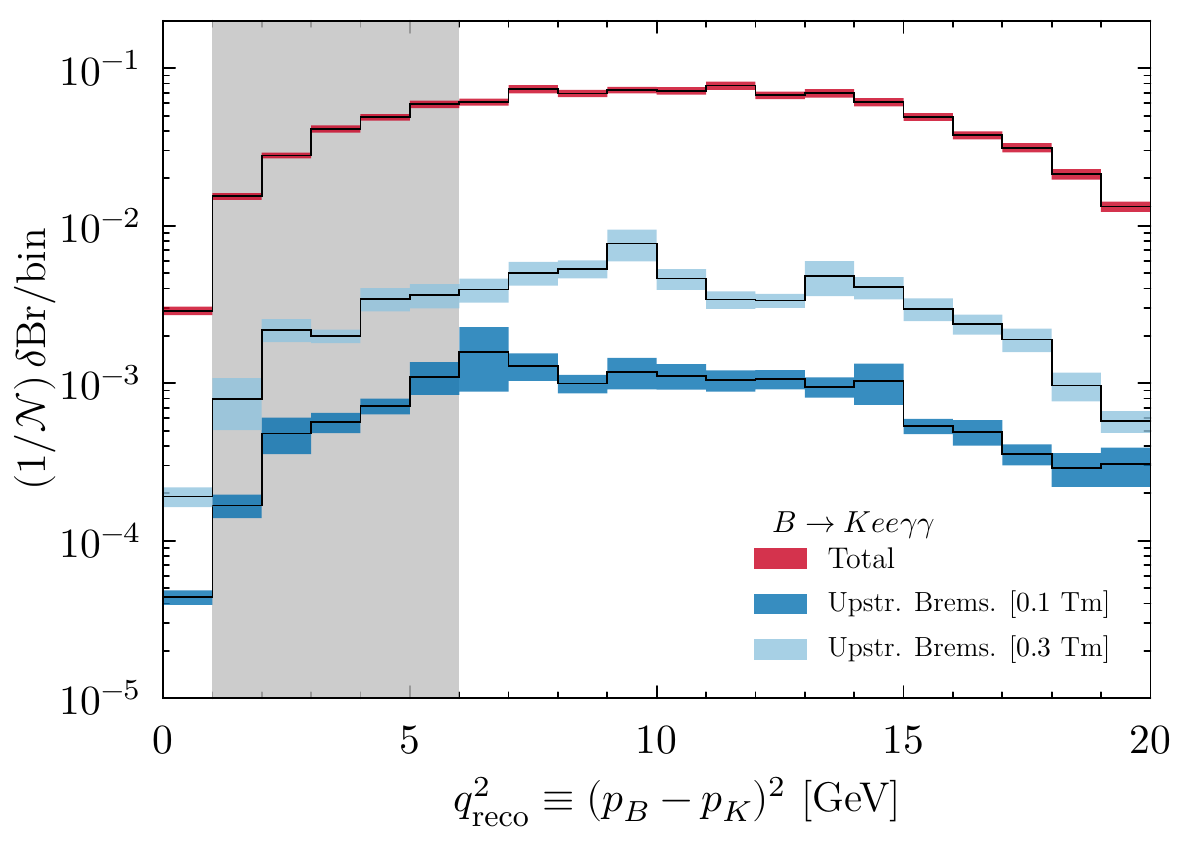}
	\caption{Normalized distributions with respect to $q^2_{\text{reco}} \equiv (p_B - p_K)^2$ from 
	the total $B \to  K (\pi^0 \to e e \gamma) \gamma$ rate (red) over the full $q^2$ range, 
	and with misreconstruction of the photons as upstream Bremsstrahlung, 
	assuming magnetic bending power $0.1$\,Tm (blue) and $0.3$\,Tm (light blue). 
	The $R_K$ signal regime $1 \le q^2_{\text{reco}} \le 6$\,GeV$^2$ is shown by the gray band.
	Uncertainties are from MC statistics alone.}
	\label{fig:q2reco}
\end{figure}

The differential angular deflection of a lepton track
\begin{equation}
	d\theta_{\text{def}} \simeq 0.3\times
	\bigg[\frac{\int B dl}{1 \text{Tm}} \bigg] \bigg[\frac{\text{GeV}}{|\bm{p}_\perp|}\bigg]\,,
\end{equation}	
over differential path length $dl$, 
with $\bm{p}_\perp$ the lepton momentum perpendicular to the magnetic field. 
Although outside the LHCb magnet the magnetic field diverges, 
I assume for simplicity that it is uniformly oriented in $y$ direction, 
with varying strength in $z$ 
(using standard beam-axis coordinates; see Fig.~\ref{fig:brems}).
From the VELO to the front face of the TT, 
the bending power is measured to be $\int B dl \simeq 0.11$\,Tm~\cite{Losasso:2006pgb}, 
increasing to approximately $0.25$\,Tm at its back face,
and approximately $0.33$\,Tm once the magnetic field has reached $0.5$\,T: half its full strength.
To be conservative, I use $\int B dl \simeq 0.1$\,Tm
in the estimate of the total angular displacement $\theta_{\text{def}}$.
In addition, I also show results for $0.3$\,Tm, 
representing a plausible scenario for the allowed upstream Bremsstrahlung emission region.

This very approximate understanding leads to the following approximate 
algorithm for simulation of a photon as Bremsstrahlung, shown in Fig~\ref{fig:brems}: 
\textbf{(i)} For each lepton track, I construct a cone of angular size $\theta_{\text{def}}$ around its truth lab frame momentum, $\bm{p}_e$;
\textbf{(ii)} Because electrons (positrons) bend in the $+x$ ($-x)$ direction, I further divide the cone in the $y$-$z$ plane and select the half-cone on the $+x$ ($-x$) side, 
corresponding to the direction of the lepton angular deflection;
\textbf{(iii)} Any photon that lies within this half-cone is considered compatible with recovery as an upstream Bremsstrahlung photon;
Finally, \textbf{(iv)}, as done in Refs.~\cite{LHCb:2014vgu,LHCb:2017avl,LHCb:2019hip,LHCb:2021trn,LHCb:2021lvy}, I require a minimum transverse momentum, $p_T$, threshold for the leptons.
The precise threshold is not provided in the recent LHCb $R_{K^+}$ analyses, however Ref.~\cite{LHCb:2017avl} specifies 
\begin{equation}
	\label{eqn:pT}
	\text{min}[p_{T}(e^+),p_{T}(e^-)] > 0.5\,\text{GeV}\,,
\end{equation}
which appears compatible with Figs. S2 of Ref.~\cite{LHCb:2019hip}.
The ECAL itself further has a finite resolution, that sets a lower bound for $\theta_{\text{def}}$. 
Details of the ECAL cell resolution are hard to glean from available literature: 
To be conservative I assume perfect ECAL resolution, 
and compare this to setting a lower bound $\theta_{\text{def}} > 5 \times 10^{-3}$,
based on the $\sim5$\,cm size of an inner or middle ECAL cell~\cite{Amato:494264,AbellanBeteta:2020amj} 
at $\sim 12$\,m from the VELO.

\section{Results}

In Fig.~\ref{fig:q2reco} I show the $q^2_{\text{reco}} \equiv (p_B - p_K)^2$ normalized distribution over the full range $q^2_{\text{reco}} \in [m_e^2, (m_B - m_K)^2]$,
generated by $B \to K (\pi^0 \to e e \gamma) \gamma$ from the above simulation.
Multiplying by the normalization factor 
\begin{equation}
	\mathcal{N} \equiv \int \! ds \frac{|F(s)|^2}{\pi}\,\text{Br}[\pi^0 \to ee \gamma] \simeq (3.0 \pm 0.2)\times 10^{-7}\,,
\end{equation} 
yields the differential branching ratio, 
including only the leading uncertainties from the $V=K^*$ contribution.
I have dropped the remaining uncertainties from the data in Table~\ref{tab:svm} that enter into $F(s)$, 
since we are interested only in an estimate, and such uncertainties
will be subleading compared to the MC uncertainties from the simulation.

Also shown in Fig.~\ref{fig:q2reco} in blue is the differential distribution
keeping only those events that satisfy the Bremsstrahlung recovery algorithm in Sec.~\ref{sec:brems}. 
The fraction of events for which both photons in $B \to K(\pi^0 \to ee\gamma)\gamma$ 
are misreconstructed as Bremsstrahlung in the $1 \le q^2_{\text{reco}} \le 6$\,GeV$^2$ signal regime
is estimated as
\begin{equation}
	\label{eqn:fmis}
	f_{\text{misreco}} =
		\begin{cases} 
			(0.31 \pm 0.08)\% &  \big[\mbox{$\int$} B dl  = 0.1\,\text{Tm}\big]\,,\\[5pt]
			(1.2 \pm 0.3)\% &  \big[\mbox{$\int$} B dl  = 0.3\,\text{Tm}\big]\,
		\end{cases}
\end{equation}
where the uncertainty is purely from MC, 
and I show results for the conservative and plausible values for the bending power,
as discussed in Sec.~\ref{sec:brems}.
Multiplying by the normalization factor $\mathcal{N}$ yields the corresponding branching ratio for the misreconstruction.

The effect of the $p_T$ threshold is significant: Without this cut, 
$f_{\text{misreco}}$ would significantly increase to $(1.7 \pm 0.4)\%$ and $(4.8 \pm 0.7)\%$,
for the conservative and plausible bending power values, respectively.
This suggests tighter $p_T$ thresholds may entirely suppress the contribution 
from misreconstructed $B \to K(\pi^0 \to ee\gamma)\gamma$ decays altogether.
Setting a lower bound $\theta_{\text{def}} > 5 \times 10^{-3}$
does not enhance the misreconstruction fraction beyond the MC uncertainties in Eq.~\eqref{eqn:fmis},
so that the ECAL resolution effects appear to be subleading.
Similarly, including the case that the $\pi^0$ daughter photon is soft, 
below a conservative $75$\,MeV threshold for Bremsstrahlung recovery~\cite{Mombacher:2020jrx},
leads to a negligible increase in $f_{\text{misreco}}$.

The measured branching ratio $\text{Br}[B^+ \to K^+ \mu\mu] = 1.2(1)\times10^{-7}$~\cite{Aaij:2014pli}. 
Taking this as a proxy for the true $B^+ \to K^+ ee$ branching fraction assuming no LFUV,
then the fractional enhancement in the measured $B^+ \to K^+ ee$ with misreconstruction is 
\begin{equation}
	\frac{\mathcal{N}f_{\text{misreco}} }{\text{Br}[B^+ \to K^+ \mu\mu]} \simeq 
	\begin{cases}
		(0.8 \pm 0.2)\% &  \big[\mbox{$\int$} B dl  = 0.1\,\text{Tm}\big]\,,\\[5pt]
		(3.1 \pm 0.8)\% &  \big[\mbox{$\int$} B dl  = 0.3\,\text{Tm}\big]\,.
	\end{cases}
\end{equation}
Thus, one roughly expects in the SM the recovered ratio could decrease to $R^{\text{rec}}_K \simeq 0.99 \pm 0.005$ and $0.97 \pm 0.01$, respectively.
The latter shift is comparable to the size of the combined statistical and systematic uncertainties quoted in Ref.~\cite{LHCb:2019hip}.

\section{Summary and outlook}
The $\pi^0 \to ee \gamma$ Dalitz decays of neutral pions produced in $b \to s \gamma$ processes  
may generate an additional sources of uncertainty in precision measurements of the LFUV ratios, $R_{K^{(*)}}$,
if the photons are misreconstructed into the $\pi^0$ daughter leptons as Bremsstrahlung.
In this note, an approximate Breit-Wigner parametrization for the resonant contributions $B \to (V \to K \pi^0)\gamma$ was
combined with a conservative guesstimate of the implementation for Bremsstrahlung recovery at LHCb,
to produce an approximate simulation of the contributions to $R_K$ from misreconstruction of $B \to K (\pi^0 \to ee \gamma)\gamma$.
This simulation suggests a LFUV background present at the percent level in $R_K$ from $B \to K \pi^0 \gamma$ alone, 
that could be as large as $3\%$, depending on the magnetic bending power.
For two cases of $\int B dl = 0.1$\,Tm and $0.3$\,Tm, 
the recovered ratio in the SM would be expected to decrease to $R^{\text{rec}}_K \simeq 0.99 \pm 0.005$
and $0.97 \pm 0.01$, respectively.
The same effect, at the same order of magnitude, may occur in $R_{K^*}$ measurements.
Other $b \to s \gamma$ transitions involving (pseudo)scalars such as $B \to K h^0\gamma$, $h^0 = \eta, \eta',\ldots$
may further enhance $R_K$, because of their enhanced Dalitz decay to $ee\gamma$ near the photon pole.
Whether modes such as $B \to K \pi^0\pi^0\gamma$ may also contribute similarly to $R_K$ requires further study.

Keeping in mind that: 
\textbf{(i)} I have likely underestimated the
 $B \to K \pi^0 \gamma$ branching ratio;
\textbf{(ii)} the effective bending power of the LHCb magnet could be even greater than the $0.3$\,Tm estimate, and;
\textbf{(iii)} that there may be other reconstruction resolution effects, that loosen the effective allowable angular displacement of a photon versus a lepton 
in order for the former to be recovered as upstream Bremsstrahlung,
it is not inconceivable that the effect on $R_{\Kx}$ could $\mathcal{O}(1)$ greater than estimated here.
Of course, it is also conceivable that:
\textbf{(iv)} the Bremsstrahlung recovery at LHCb could be far better able to discriminate or reject fakes than estimated above, and;
\textbf{(v)} the reconstruction efficiency for the $B \to K (\pi^0 \to ee \gamma)\gamma$ background could be much lower than for the signal, 
leading to a substantial suppression of the effect on $R_{\Kx}$.
A proper estimate of this effect will require dedicated studies within experimental frameworks, not only at LHCb but also at Belle~II.

If the effect is present, then precision measurements of the LFUV ratios $R_{\Kx}$ 
will require improved theoretical descriptions of $B \to K\pi^0\gamma$-like backgrounds, which feature notable hadronic uncertainties.
If such a scenario arises, this will be (yet another) example of a theoretically clean observable that acquires nontrivial theoretical uncertainties 
when recovered from a precision experimental framework.

\acknowledgements
I thank Florian Bernlochner, Marat Freytsis, Zoltan Ligeti, Michele Papucci and Maayan Robinson for discussions and for their comments on the manuscript.
I further thank Marat Freytsis for pointing out the possibility of contributions from other modes such as $B \to K \pi^0\pi^0\gamma$,
and I thank the Referees for their comments and observations, in particular pointing out the importance of $p_T$ selections for the leptons.
This work is supported by the Office of High Energy Physics of the U.S. Department of Energy under contract DE-AC02-05CH11231.


\begin{thebibliography}{25}%
\makeatletter
\providecommand \@ifxundefined [1]{%
 \@ifx{#1\undefined}
}%
\providecommand \@ifnum [1]{%
 \ifnum #1\expandafter \@firstoftwo
 \else \expandafter \@secondoftwo
 \fi
}%
\providecommand \@ifx [1]{%
 \ifx #1\expandafter \@firstoftwo
 \else \expandafter \@secondoftwo
 \fi
}%
\providecommand \natexlab [1]{#1}%
\providecommand \enquote  [1]{``#1''}%
\providecommand \bibnamefont  [1]{#1}%
\providecommand \bibfnamefont [1]{#1}%
\providecommand \citenamefont [1]{#1}%
\providecommand \href@noop [0]{\@secondoftwo}%
\providecommand \href [0]{\begingroup \@sanitize@url \@href}%
\providecommand \@href[1]{\@@startlink{#1}\@@href}%
\providecommand \@@href[1]{\endgroup#1\@@endlink}%
\providecommand \@sanitize@url [0]{\catcode `\\12\catcode `\$12\catcode
  `\&12\catcode `\#12\catcode `\^12\catcode `\_12\catcode `\%12\relax}%
\providecommand \@@startlink[1]{}%
\providecommand \@@endlink[0]{}%
\providecommand \url  [0]{\begingroup\@sanitize@url \@url }%
\providecommand \@url [1]{\endgroup\@href {#1}{\urlprefix }}%
\providecommand \urlprefix  [0]{URL }%
\providecommand \Eprint [0]{\href }%
\providecommand \doibase [0]{http://dx.doi.org/}%
\providecommand \selectlanguage [0]{\@gobble}%
\providecommand \bibinfo  [0]{\@secondoftwo}%
\providecommand \bibfield  [0]{\@secondoftwo}%
\providecommand \translation [1]{[#1]}%
\providecommand \BibitemOpen [0]{}%
\providecommand \bibitemStop [0]{}%
\providecommand \bibitemNoStop [0]{.\EOS\space}%
\providecommand \EOS [0]{\spacefactor3000\relax}%
\providecommand \BibitemShut  [1]{\csname bibitem#1\endcsname}%
\let\auto@bib@innerbib\@empty
\bibitem [{\citenamefont {Bordone}\ \emph {et~al.}(2016)\citenamefont
  {Bordone}, \citenamefont {Isidori},\ and\ \citenamefont
  {Pattori}}]{Bordone:2016gaq}%
  \BibitemOpen
  \bibfield  {author} {\bibinfo {author} {\bibfnamefont {M.}~\bibnamefont
  {Bordone}}, \bibinfo {author} {\bibfnamefont {G.}~\bibnamefont {Isidori}}, \
  and\ \bibinfo {author} {\bibfnamefont {A.}~\bibnamefont {Pattori}},\ }\href
  {\doibase 10.1140/epjc/s10052-016-4274-7} {\bibfield  {journal} {\bibinfo
  {journal} {Eur. Phys. J. C}\ }\textbf {\bibinfo {volume} {76}},\ \bibinfo
  {pages} {440} (\bibinfo {year} {2016})},\ \Eprint
  {http://arxiv.org/abs/1605.07633} {arXiv:1605.07633 [hep-ph]} \BibitemShut
  {NoStop}%
\bibitem [{\citenamefont {Zyla}\ \emph {et~al.}(2020)\citenamefont {Zyla} \emph
  {et~al.}}]{Zyla:2020zbs}%
  \BibitemOpen
  \bibfield  {author} {\bibinfo {author} {\bibfnamefont {P.}~\bibnamefont
  {Zyla}} \emph {et~al.} (\bibinfo {collaboration} {Particle Data Group}),\
  }\href {\doibase 10.1093/ptep/ptaa104} {\bibfield  {journal} {\bibinfo
  {journal} {PTEP}\ }\textbf {\bibinfo {volume} {2020}},\ \bibinfo {pages}
  {083C01} (\bibinfo {year} {2020})}\BibitemShut {NoStop}%
\bibitem [{\citenamefont {Aaij}\ \emph
  {et~al.}(2021{\natexlab{a}})\citenamefont {Aaij} \emph
  {et~al.}}]{LHCb:2021trn}%
  \BibitemOpen
  \bibfield  {author} {\bibinfo {author} {\bibfnamefont {R.}~\bibnamefont
  {Aaij}} \emph {et~al.} (\bibinfo {collaboration} {LHCb}),\ }\href@noop {} {\
  (\bibinfo {year} {2021}{\natexlab{a}})},\ \Eprint
  {http://arxiv.org/abs/2103.11769} {arXiv:2103.11769 [hep-ex]} \BibitemShut
  {NoStop}%
\bibitem [{\citenamefont {Aaij}\ \emph {et~al.}(2019)\citenamefont {Aaij} \emph
  {et~al.}}]{LHCb:2019hip}%
  \BibitemOpen
  \bibfield  {author} {\bibinfo {author} {\bibfnamefont {R.}~\bibnamefont
  {Aaij}} \emph {et~al.} (\bibinfo {collaboration} {LHCb}),\ }\href {\doibase
  10.1103/PhysRevLett.122.191801} {\bibfield  {journal} {\bibinfo  {journal}
  {Phys. Rev. Lett.}\ }\textbf {\bibinfo {volume} {122}},\ \bibinfo {pages}
  {191801} (\bibinfo {year} {2019})},\ \Eprint
  {http://arxiv.org/abs/1903.09252} {arXiv:1903.09252 [hep-ex]} \BibitemShut
  {NoStop}%
\bibitem [{\citenamefont {Aaij}\ \emph
  {et~al.}(2021{\natexlab{b}})\citenamefont {Aaij} \emph
  {et~al.}}]{LHCb:2021lvy}%
  \BibitemOpen
  \bibfield  {author} {\bibinfo {author} {\bibfnamefont {R.}~\bibnamefont
  {Aaij}} \emph {et~al.} (\bibinfo {collaboration} {LHCb}),\ }\href@noop {} {\
  (\bibinfo {year} {2021}{\natexlab{b}})},\ \Eprint
  {http://arxiv.org/abs/2110.09501} {arXiv:2110.09501 [hep-ex]} \BibitemShut
  {NoStop}%
\bibitem [{\citenamefont {Bobeth}\ \emph {et~al.}(2007)\citenamefont {Bobeth},
  \citenamefont {Hiller},\ and\ \citenamefont {Piranishvili}}]{Bobeth:2007dw}%
  \BibitemOpen
  \bibfield  {author} {\bibinfo {author} {\bibfnamefont {C.}~\bibnamefont
  {Bobeth}}, \bibinfo {author} {\bibfnamefont {G.}~\bibnamefont {Hiller}}, \
  and\ \bibinfo {author} {\bibfnamefont {G.}~\bibnamefont {Piranishvili}},\
  }\href {\doibase 10.1088/1126-6708/2007/12/040} {\bibfield  {journal}
  {\bibinfo  {journal} {JHEP}\ }\textbf {\bibinfo {volume} {12}},\ \bibinfo
  {pages} {040} (\bibinfo {year} {2007})},\ \Eprint
  {http://arxiv.org/abs/0709.4174} {arXiv:0709.4174 [hep-ph]} \BibitemShut
  {NoStop}%
\bibitem [{\citenamefont {Descotes-Genon}\ \emph {et~al.}(2016)\citenamefont
  {Descotes-Genon}, \citenamefont {Hofer}, \citenamefont {Matias},\ and\
  \citenamefont {Virto}}]{Descotes-Genon:2015uva}%
  \BibitemOpen
  \bibfield  {author} {\bibinfo {author} {\bibfnamefont {S.}~\bibnamefont
  {Descotes-Genon}}, \bibinfo {author} {\bibfnamefont {L.}~\bibnamefont
  {Hofer}}, \bibinfo {author} {\bibfnamefont {J.}~\bibnamefont {Matias}}, \
  and\ \bibinfo {author} {\bibfnamefont {J.}~\bibnamefont {Virto}},\ }\href
  {\doibase 10.1007/JHEP06(2016)092} {\bibfield  {journal} {\bibinfo  {journal}
  {JHEP}\ }\textbf {\bibinfo {volume} {06}},\ \bibinfo {pages} {092} (\bibinfo
  {year} {2016})},\ \Eprint {http://arxiv.org/abs/1510.04239} {arXiv:1510.04239
  [hep-ph]} \BibitemShut {NoStop}%
\bibitem [{\citenamefont {Isidori}\ \emph {et~al.}(2020)\citenamefont
  {Isidori}, \citenamefont {Nabeebaccus},\ and\ \citenamefont
  {Zwicky}}]{Isidori:2020acz}%
  \BibitemOpen
  \bibfield  {author} {\bibinfo {author} {\bibfnamefont {G.}~\bibnamefont
  {Isidori}}, \bibinfo {author} {\bibfnamefont {S.}~\bibnamefont
  {Nabeebaccus}}, \ and\ \bibinfo {author} {\bibfnamefont {R.}~\bibnamefont
  {Zwicky}},\ }\href {\doibase 10.1007/JHEP12(2020)104} {\bibfield  {journal}
  {\bibinfo  {journal} {JHEP}\ }\textbf {\bibinfo {volume} {12}},\ \bibinfo
  {pages} {104} (\bibinfo {year} {2020})},\ \Eprint
  {http://arxiv.org/abs/2009.00929} {arXiv:2009.00929 [hep-ph]} \BibitemShut
  {NoStop}%
\bibitem [{\citenamefont {Aaij}\ \emph
  {et~al.}(2014{\natexlab{a}})\citenamefont {Aaij} \emph
  {et~al.}}]{LHCb:2014vgu}%
  \BibitemOpen
  \bibfield  {author} {\bibinfo {author} {\bibfnamefont {R.}~\bibnamefont
  {Aaij}} \emph {et~al.} (\bibinfo {collaboration} {LHCb}),\ }\href {\doibase
  10.1103/PhysRevLett.113.151601} {\bibfield  {journal} {\bibinfo  {journal}
  {Phys. Rev. Lett.}\ }\textbf {\bibinfo {volume} {113}},\ \bibinfo {pages}
  {151601} (\bibinfo {year} {2014}{\natexlab{a}})},\ \Eprint
  {http://arxiv.org/abs/1406.6482} {arXiv:1406.6482 [hep-ex]} \BibitemShut
  {NoStop}%
\bibitem [{\citenamefont {Aaij}\ \emph {et~al.}(2017)\citenamefont {Aaij} \emph
  {et~al.}}]{LHCb:2017avl}%
  \BibitemOpen
  \bibfield  {author} {\bibinfo {author} {\bibfnamefont {R.}~\bibnamefont
  {Aaij}} \emph {et~al.} (\bibinfo {collaboration} {LHCb}),\ }\href {\doibase
  10.1007/JHEP08(2017)055} {\bibfield  {journal} {\bibinfo  {journal} {JHEP}\
  }\textbf {\bibinfo {volume} {08}},\ \bibinfo {pages} {055} (\bibinfo {year}
  {2017})},\ \Eprint {http://arxiv.org/abs/1705.05802} {arXiv:1705.05802
  [hep-ex]} \BibitemShut {NoStop}%
\bibitem [{\citenamefont {Wei}\ \emph {et~al.}(2009)\citenamefont {Wei} \emph
  {et~al.}}]{Belle:2009zue}%
  \BibitemOpen
  \bibfield  {author} {\bibinfo {author} {\bibfnamefont {J.~T.}\ \bibnamefont
  {Wei}} \emph {et~al.} (\bibinfo {collaboration} {Belle}),\ }\href {\doibase
  10.1103/PhysRevLett.103.171801} {\bibfield  {journal} {\bibinfo  {journal}
  {Phys. Rev. Lett.}\ }\textbf {\bibinfo {volume} {103}},\ \bibinfo {pages}
  {171801} (\bibinfo {year} {2009})},\ \Eprint {http://arxiv.org/abs/0904.0770}
  {arXiv:0904.0770 [hep-ex]} \BibitemShut {NoStop}%
\bibitem [{\citenamefont {Abdesselam}\ \emph {et~al.}(2021)\citenamefont
  {Abdesselam} \emph {et~al.}}]{Belle:2019oag}%
  \BibitemOpen
  \bibfield  {author} {\bibinfo {author} {\bibfnamefont {A.}~\bibnamefont
  {Abdesselam}} \emph {et~al.} (\bibinfo {collaboration} {Belle}),\ }\href
  {\doibase 10.1103/PhysRevLett.126.161801} {\bibfield  {journal} {\bibinfo
  {journal} {Phys. Rev. Lett.}\ }\textbf {\bibinfo {volume} {126}},\ \bibinfo
  {pages} {161801} (\bibinfo {year} {2021})},\ \Eprint
  {http://arxiv.org/abs/1904.02440} {arXiv:1904.02440 [hep-ex]} \BibitemShut
  {NoStop}%
\bibitem [{\citenamefont {Choudhury}\ \emph {et~al.}(2021)\citenamefont
  {Choudhury} \emph {et~al.}}]{BELLE:2019xld}%
  \BibitemOpen
  \bibfield  {author} {\bibinfo {author} {\bibfnamefont {S.}~\bibnamefont
  {Choudhury}} \emph {et~al.} (\bibinfo {collaboration} {BELLE}),\ }\href
  {\doibase 10.1007/JHEP03(2021)105} {\bibfield  {journal} {\bibinfo  {journal}
  {JHEP}\ }\textbf {\bibinfo {volume} {03}},\ \bibinfo {pages} {105} (\bibinfo
  {year} {2021})},\ \Eprint {http://arxiv.org/abs/1908.01848} {arXiv:1908.01848
  [hep-ex]} \BibitemShut {NoStop}%
\bibitem [{\citenamefont {Aubert}\ \emph {et~al.}(2009)\citenamefont {Aubert}
  \emph {et~al.}}]{BaBar:2008jdv}%
  \BibitemOpen
  \bibfield  {author} {\bibinfo {author} {\bibfnamefont {B.}~\bibnamefont
  {Aubert}} \emph {et~al.} (\bibinfo {collaboration} {BaBar}),\ }\href
  {\doibase 10.1103/PhysRevLett.102.091803} {\bibfield  {journal} {\bibinfo
  {journal} {Phys. Rev. Lett.}\ }\textbf {\bibinfo {volume} {102}},\ \bibinfo
  {pages} {091803} (\bibinfo {year} {2009})},\ \Eprint
  {http://arxiv.org/abs/0807.4119} {arXiv:0807.4119 [hep-ex]} \BibitemShut
  {NoStop}%
\bibitem [{\citenamefont {Aaij}\ \emph
  {et~al.}(2021{\natexlab{c}})\citenamefont {Aaij} \emph
  {et~al.}}]{LHCb:2021rou}%
  \BibitemOpen
  \bibfield  {author} {\bibinfo {author} {\bibfnamefont {R.}~\bibnamefont
  {Aaij}} \emph {et~al.} (\bibinfo {collaboration} {LHCb}),\ }\href {\doibase
  10.1007/JHEP06(2021)019} {\bibfield  {journal} {\bibinfo  {journal} {JHEP}\
  }\textbf {\bibinfo {volume} {06}},\ \bibinfo {pages} {019} (\bibinfo {year}
  {2021}{\natexlab{c}})},\ \Eprint {http://arxiv.org/abs/2103.11058}
  {arXiv:2103.11058 [hep-ex]} \BibitemShut {NoStop}%
\bibitem [{\citenamefont {Calvo~Gomez}\ \emph {et~al.}(2015)\citenamefont
  {Calvo~Gomez}, \citenamefont {Cogneras}, \citenamefont {Deschamps},
  \citenamefont {Hoballah}, \citenamefont {Lefevre}, \citenamefont {Monteil},
  \citenamefont {Puig~Navarro},\ and\ \citenamefont
  {Rives~Molina}}]{CalvoGomez:2042173}%
  \BibitemOpen
  \bibfield  {author} {\bibinfo {author} {\bibfnamefont {M.}~\bibnamefont
  {Calvo~Gomez}}, \bibinfo {author} {\bibfnamefont {E.}~\bibnamefont
  {Cogneras}}, \bibinfo {author} {\bibfnamefont {O.}~\bibnamefont {Deschamps}},
  \bibinfo {author} {\bibfnamefont {M.}~\bibnamefont {Hoballah}}, \bibinfo
  {author} {\bibfnamefont {R.}~\bibnamefont {Lefevre}}, \bibinfo {author}
  {\bibfnamefont {S.}~\bibnamefont {Monteil}}, \bibinfo {author} {\bibfnamefont
  {A.}~\bibnamefont {Puig~Navarro}}, \ and\ \bibinfo {author} {\bibfnamefont
  {V.~J.}\ \bibnamefont {Rives~Molina}},\ }\href
  {https://cds.cern.ch/record/2042173} {\emph {\bibinfo {title} {{A tool for
  $\gamma/\pi^0$ separation at high energies}}}},\ \bibinfo {type} {Tech.
  Rep.}\ \bibinfo {number} {LHCb-PUB-2015-016, CERN-LHCb-PUB-2015-016}\
  (\bibinfo  {institution} {CERN},\ \bibinfo {address} {Geneva},\ \bibinfo
  {year} {2015})\BibitemShut {NoStop}%
\bibitem [{\citenamefont {Aaij}\ \emph {et~al.}(2016)\citenamefont {Aaij} \emph
  {et~al.}}]{Aaij:2016mos}%
  \BibitemOpen
  \bibfield  {author} {\bibinfo {author} {\bibfnamefont {R.}~\bibnamefont
  {Aaij}} \emph {et~al.} (\bibinfo {collaboration} {LHCb}),\ }\href {\doibase
  10.1103/PhysRevLett.116.241601} {\bibfield  {journal} {\bibinfo  {journal}
  {Phys. Rev. Lett.}\ }\textbf {\bibinfo {volume} {116}},\ \bibinfo {pages}
  {241601} (\bibinfo {year} {2016})},\ \Eprint
  {http://arxiv.org/abs/1603.04804} {arXiv:1603.04804 [hep-ex]} \BibitemShut
  {NoStop}%
\bibitem [{\citenamefont {Aguil{\'o}}\ \emph {et~al.}(2006)\citenamefont
  {Aguil{\'o}}, \citenamefont {Calvo},\ and\ \citenamefont
  {Garrido}}]{Aguilo:1000431}%
  \BibitemOpen
  \bibfield  {author} {\bibinfo {author} {\bibfnamefont {E.}~\bibnamefont
  {Aguil{\'o}}}, \bibinfo {author} {\bibfnamefont {M.}~\bibnamefont {Calvo}}, \
  and\ \bibinfo {author} {\bibfnamefont {L.}~\bibnamefont {Garrido}},\ }\href
  {https://cds.cern.ch/record/1000431} {\emph {\bibinfo {title} {{Recovery of
  radiated energy in the $B_d \to J/\psi(e^+e^-)K_S$ decay channel}}}},\
  \bibinfo {type} {Tech. Rep.}\ \bibinfo {number} {{LHCb-2006-062,
  CERN-LHCb-2006-062}}\ (\bibinfo  {institution} {CERN},\ \bibinfo {address}
  {Geneva},\ \bibinfo {year} {2006})\BibitemShut {NoStop}%
\bibitem [{\citenamefont {Momb\"acher}(2020)}]{Mombacher:2020jrx}%
  \BibitemOpen
  \bibfield  {author} {\bibinfo {author} {\bibfnamefont {T.}~\bibnamefont
  {Momb\"acher}},\ }\emph {\bibinfo {title} {{Beautiful leptons - setting
  limits to New Physics with the LHCb experiment}}},\ \href {\doibase
  10.17877/DE290R-21750} {Ph.D. thesis},\ \bibinfo  {school} {Tech. U.,
  Dortmund (main)} (\bibinfo {year} {2020}),\ \bibinfo {note}
  {{CERN-THESIS-2020-246}}\BibitemShut {NoStop}%
\bibitem [{\citenamefont {Berninghoff}\ \emph {et~al.}(2016)\citenamefont
  {Berninghoff}, \citenamefont {Albrecht},\ and\ \citenamefont
  {Gligorov}}]{Berninghoff:2146447}%
  \BibitemOpen
  \bibfield  {author} {\bibinfo {author} {\bibfnamefont {D.~A.}\ \bibnamefont
  {Berninghoff}}, \bibinfo {author} {\bibfnamefont {J.}~\bibnamefont
  {Albrecht}}, \ and\ \bibinfo {author} {\bibfnamefont {V.}~\bibnamefont
  {Gligorov}},\ }\href {https://cds.cern.ch/record/2146447} {\emph {\bibinfo
  {title} {{Bremsstrahlung Recovery of Electrons using Multivariate
  Methods}}}},\ \bibinfo {type} {Tech. Rep.}\ \bibinfo {number}
  {LHCb-INT-2016-018, CERN-LHCb-INT-2016-018}\ (\bibinfo  {institution}
  {CERN},\ \bibinfo {address} {Geneva},\ \bibinfo {year} {2016})\BibitemShut
  {NoStop}%
\bibitem [{\citenamefont {Terrier}\ and\ \citenamefont
  {Belyaev}(2003)}]{Terrier:691743}%
  \BibitemOpen
  \bibfield  {author} {\bibinfo {author} {\bibfnamefont {H.}~\bibnamefont
  {Terrier}}\ and\ \bibinfo {author} {\bibfnamefont {I.}~\bibnamefont
  {Belyaev}},\ }\href {https://cds.cern.ch/record/691743} {\emph {\bibinfo
  {title} {{Particle identification with LHCb calorimeters}}}},\ \bibinfo
  {type} {Tech. Rep.}\ \bibinfo {number} {LHCb-2003-092}\ (\bibinfo
  {institution} {CERN},\ \bibinfo {address} {Geneva},\ \bibinfo {year}
  {2003})\BibitemShut {NoStop}%
\bibitem [{\citenamefont {Losasso}\ \emph {et~al.}(2006)\citenamefont
  {Losasso}, \citenamefont {Bersgma}, \citenamefont {Flegel}, \citenamefont
  {Giudici}, \citenamefont {Hernando}, \citenamefont {Jamet}, \citenamefont
  {Lindner}, \citenamefont {Renaud},\ and\ \citenamefont
  {Teubert}}]{Losasso:2006pgb}%
  \BibitemOpen
  \bibfield  {author} {\bibinfo {author} {\bibfnamefont {M.}~\bibnamefont
  {Losasso}}, \bibinfo {author} {\bibfnamefont {F.}~\bibnamefont {Bersgma}},
  \bibinfo {author} {\bibfnamefont {W.}~\bibnamefont {Flegel}}, \bibinfo
  {author} {\bibfnamefont {P.~A.}\ \bibnamefont {Giudici}}, \bibinfo {author}
  {\bibfnamefont {J.~A.}\ \bibnamefont {Hernando}}, \bibinfo {author}
  {\bibfnamefont {O.}~\bibnamefont {Jamet}}, \bibinfo {author} {\bibfnamefont
  {R.}~\bibnamefont {Lindner}}, \bibinfo {author} {\bibfnamefont
  {J.}~\bibnamefont {Renaud}}, \ and\ \bibinfo {author} {\bibfnamefont
  {F.}~\bibnamefont {Teubert}},\ }\href {\doibase 10.1109/TASC.2005.869655}
  {\bibfield  {journal} {\bibinfo  {journal} {IEEE Trans. Appl. Supercond.}\
  }\textbf {\bibinfo {volume} {16}},\ \bibinfo {pages} {1700} (\bibinfo {year}
  {2006})},\ \bibinfo {note} {{LHCb-PROC-2005-028}}\BibitemShut {NoStop}%
\bibitem [{\citenamefont {Amato}\ \emph {et~al.}(2000)\citenamefont {Amato}
  \emph {et~al.}}]{Amato:494264}%
  \BibitemOpen
  \bibfield  {author} {\bibinfo {author} {\bibfnamefont {S.}~\bibnamefont
  {Amato}} \emph {et~al.} (\bibinfo {collaboration} {LHCb Collaboration}),\
  }\href {https://cds.cern.ch/record/494264} {\emph {\bibinfo {title} {{LHCb
  calorimeters: Technical Design Report}}}},\ \bibinfo {number}
  {CERN-LHCC-2000-036, LHCb-TDR-2}\ (\bibinfo  {publisher} {CERN},\ \bibinfo
  {address} {Geneva},\ \bibinfo {year} {2000})\BibitemShut {NoStop}%
\bibitem [{\citenamefont {Abell\'an~Beteta}\ \emph {et~al.}(2020)\citenamefont
  {Abell\'an~Beteta} \emph {et~al.}}]{AbellanBeteta:2020amj}%
  \BibitemOpen
  \bibfield  {author} {\bibinfo {author} {\bibfnamefont {C.}~\bibnamefont
  {Abell\'an~Beteta}} \emph {et~al.},\ }\href@noop {} {\  (\bibinfo {year}
  {2020})},\ \bibinfo {note} {{CERN-LHCb-DP-2020-001}},\ \Eprint
  {http://arxiv.org/abs/2008.11556} {arXiv:2008.11556 [physics.ins-det]}
  \BibitemShut {NoStop}%
\bibitem [{\citenamefont {Aaij}\ \emph
  {et~al.}(2014{\natexlab{b}})\citenamefont {Aaij} \emph
  {et~al.}}]{Aaij:2014pli}%
  \BibitemOpen
  \bibfield  {author} {\bibinfo {author} {\bibfnamefont {R.}~\bibnamefont
  {Aaij}} \emph {et~al.} (\bibinfo {collaboration} {LHCb}),\ }\href {\doibase
  10.1007/JHEP06(2014)133} {\bibfield  {journal} {\bibinfo  {journal} {JHEP}\
  }\textbf {\bibinfo {volume} {06}},\ \bibinfo {pages} {133} (\bibinfo {year}
  {2014}{\natexlab{b}})},\ \Eprint {http://arxiv.org/abs/1403.8044}
  {arXiv:1403.8044 [hep-ex]} \BibitemShut {NoStop}%
\end{thebibliography}
\end{document}